\documentclass[aps,prl,twocolumn,showpacs]{revtex4}
\usepackage{epsfig}
\usepackage{graphicx}
\usepackage{amsfonts}
\usepackage[figuresright]{rotating} 
\usepackage{amssymb}
\usepackage{amsmath}

\def\avg#1{\langle#1\rangle}

\def\be{\begin{equation}} \def\ee{\end{equation}}
\def\bea{\begin{eqnarray}} \def\eea{\end{eqnarray}}

\def\pp{\parallel}

\begin{document}

\title{Orbital ordering and frustration of $p$-band Mott-insulators}
\author{Congjun Wu}
\affiliation{Department of Physics, University of California, San Diego,
CA 92093}

\begin{abstract}
We investigate the general structure of orbital exchange physics in 
Mott-insulating states of $p$-orbital systems in optical lattices.
Orbital orders occur in both the triangular and Kagome lattices.
In contrast, orbital exchange in the honeycomb lattice is
frustrated as described by a novel quantum 120$^\circ$-model.
Its classical ground states are mapped into  configurations
of the fully-packed loop model with an extra $U(1)$ rotation 
degree of freedom.
Quantum orbital fluctuations select a six-site plaquette ground state ordering
pattern in the semiclassical limit from the ``order from disorder'' mechanism.
This effect arises from the appearance of a zero energy flat-band 
of orbital excitations.
\end{abstract}
\maketitle

Orbital is a degree of freedom characterized by orbital degeneracy 
and orientational anisotropy.
The interplay among orbital, spin and charge degrees of freedom
gives rise to important effects on  metal-insulator transitions, 
superconductivity, and colossal magneto-resistance
in transition metal oxides \cite{Imada1998, tokura2000, khaliullin2005}.
The progress of cold atom physics in optical lattices
has provide a new opportunity to investigate orbital physics.
A major advantage of optical lattices is the absence of the Jahn-Teller
lattice distortion which lifts orbital degeneracy and
quenches the orbital degree of freedom in solid state systems.

Orbital physics in optical lattices exhibits different features 
from those in solid state systems
\cite{scarola2005, isacsson2005, liu2006, kuklov2006, wu2006,
xu2007, xu2007a, wu2007, alon2005, browaeys2005, kohl2005,
sebby-strabley2006, mueller2007}.
The Hubbard interaction of $p$-orbital bosons has the 
ferro-orbital nature leading to an ``orbital Hund's rule''.
This generates a class of orbital superfluid states with complex-valued 
wavefunctions breaking time reversal symmetry beyond Feynman's 
celebrated argument of the positive-definitive ground state 
wavefunctions  \cite{liu2006, kuklov2006, wu2006}.
The $p$-orbital honeycomb lattice filled with fermions provides a 
$p_{x,y}$-orbital counterpart of graphene, whose flat band 
structure dramatically enhances interaction effects and gives rise 
to various charge and bond crystalline orders \cite{wu2007,wu2007a}.
The experiment progress is truly exciting \cite{browaeys2005, 
kohl2005, sebby-strabley2006, mueller2007}.
In particular, the meta-stable $p$-orbital bosonic systems have been
realized by using the stimulated Raman transition to pump bosons 
into high orbital bands \cite{mueller2007}.

In Mott-insulators, the orbital degree of freedom also enables
super-exchange interaction just as spin does.
A marked difference between orbital and spin exchanges is that
the former depends on bond orientation.
Orbital exchange physics has been extensively investigated in the 
$d$-orbital $t_{2g}$ and $e_g$ systems \cite{khaliullin2005,vandenbrink2004,
biskup2005,nussinov2004,diep2004}.
However, correlation effects in the $p$-orbital bands in solid state 
systems are typically weak.
To our knowledge, the $p$-orbital exchange physics has not been 
investigated in solid state systems.
In contrast, the $p$-orbital systems in optical lattices 
can be easily tuned to the strong correlation regime, providing an 
opportunity to investigate new orbital physics.
A discussion of the $p$-orbital exchange in the honeycomb
lattice and the consequential 120$^\circ$ degree model
was presented by the author in Ref. \cite{wu2007v1}.

In this article, we construct the general structure of the $p$-orbital
exchange models in optical lattices. % whose spatial anisotropy is 
%even stronger than those of the $d$-orbitals.
Orbital orders are found in the square lattice, 
and also in the triangular and Kagome lattices which are typical 
frustrated lattices for spin systems.
In contrast, strong orbital frustration occurs in the honeycomb lattice
as described by a novel 120$^\circ$-orbital exchange model.
The classical ground states are closely related to the fully-packed 
loop representation of the three-coloring model.
The ``order from disorder'' mechanism generates a plaquette orbital 
ordering pattern form quantum orbital fluctuations.

We begin with the two dimensional $p_{x,y}$-orbital Mott-insulators
with spinless fermions by loading a single component of fermion atoms.
Each optical site is approximated by an anisotropic harmonic 
potential well with frequencies $\omega_z\gg\omega_x=\omega_y$.
Suppose that the filling is two fermions per site:
one is in the inert $s$-orbital and the rest fills the
$p_{x,y}$-orbitals.
The hopping terms in the $p$-bands can be classified as the 
$\sigma$-bonding $t_\pp$ and $\pi$-bonding $t_\perp$
(typically $t_\pp/t_\perp\gg 1$).
Due to the orbital degeneracy, the on-site interaction for spinless fermions
is still the Hubbard-like as
$H_{int}= U \sum_{\vec r } n_{\vec r, x} n_{\vec r, y}$.
For spinless fermions, the leading contribution to $U$ is from
the $p$-wave scattering.
In order to enhance $U$, we suggest using fermions with large magnetic 
moments polarized in external magnetic field, such as $^{53}$Cr with $6\mu_B$.
%The offsite interactions decay fast as $1/r^3$. 
%As estimated in Ref. \cite{wu2007},  the nearest neighbor interaction  
%$U^\prime$ can be easily tuned two orders smaller than $U$, thus 
%can be safely neglected. 
Another method is to use the $p$-wave Feshbach resonance to enhance $U
\gg t_\pp$.
But we do not need very close to the resonance, so that $U$ is still
smaller than the gap between $s$ and $p$-bands to avoid multi-band effect.
%This is helpful to maintain the system stability which is also
%benefited  by the fact that the average fermion number per site is two.
%In the following, we assume $U$ is much larger the band width but is
%still smaller than the gap between the $s$ and $p$-bands,
%thus the hybridization between $s$ and $p$-bands is negligible.

The $p$-orbital exchange physics can be conveniently represented
by using the pseudospin $\tau$-vectors defined as
$\tau_1= \frac{1}{2} (p^\dagger_x p_x-p^\dagger_y p_y), 
\tau_2 = \frac{1}{2} (p^\dagger_x p_y+h.c.),  %\nn \\
\tau_3 = \frac{1}{2i} (p^\dagger_x p_y-h.c.)$,
where $\tau_{1,2}$ describe the preferential occupation of 
orbital orientation, $\tau_3$ is the orbital angular momentum.
Let us first look at the $x$-bond.
%The four possible combinations of $p_{x,y}$ on  two neighboring sites
%are depicted in Fig. \ref{fig:exchange}.
%The configurations of A) and B) have zero exchange interaction due to
%Pauli's exclusion principle, and those of C) and D) gain the
%exchange energy of $t_\pp^2/U$ but without orbital-flip
%process in the absence of $\pi$-bonding.
Since $p_{x,y}$-orbitals are eigenstates of $\tau_1$ with eigenvalues of 
$\pm \frac{1}{2}$, respectively, this exchange is Ising-like 
in the absence of $\pi$-bonding as
$H_{ex}  (\vec r, \vec r^\prime) 
= J_\pp \tau_1 (\vec r) \tau_1(\vec r^\prime)$ with $J_\pp=2t_\pp^2/U$.
%The $t_\perp$ term can bring a small orbital flip terms as
%$J_\perp \sum_{i=2,3} \tau_i (\vec r) \tau_i(\vec r^\prime)$ with
%$J_\perp = 2t_\pp t_\perp/U$, which will be neglected below.
Moreover, the Ising quantization axis changes with bond orientations.
For a bond along a general direction of $\hat e_\varphi=\cos \varphi
\hat e_x +\sin \varphi \hat e_y$, we can rotate $p_{x,y}$-orbitals 
at an angle of $\varphi$.
In the new basis of $p^\prime_x=\cos \varphi p_x +\sin \varphi p_y$ and $
p^\prime_y=-\sin \varphi p_x+\cos \varphi p_y$ which are
the eigenstates of $\cos 2\varphi \tau_1+\sin 2\varphi 
\tau_2$, the exchange along the bond remains Ising-like as 
\bea
H_{ex}(\vec r, \vec r+\hat e_\varphi)=J_\pp
[\vec \tau (\vec r) \cdot \hat e_{2\varphi}]
[\vec \tau (\vec r+ \hat e_\varphi) \cdot \hat e_{2\varphi}].
\label{eq:exchange}
\eea
The exchange model in the lattice is just a summation of Eq. \ref{eq:exchange} 
over all the bonds.
Although the $p$-orbital system is of pseudospin-1/2, we will  
take the $\tau$-operators as general spin-$S$ operators below.
% and investigate the
%large $S$-limit.

%\begin{figure}
%\centering\epsfig{file=exchange.eps,clip=1,width=0.55\linewidth,angle=0}
%\caption{Four orbital configurations along a $x$-bond.
%The $p$-orbital exchange is Ising-like with the Ising quantization
%axis correlated with bond orientation.
%Configurations of $A$ and $B$ have zero exchange while those
%of $C$ and $D$ gain the exchange energy of $t_\pp^2/U$.
%}\label{fig:exchange}
%\end{figure}

\begin{figure}
\centering\epsfig{file=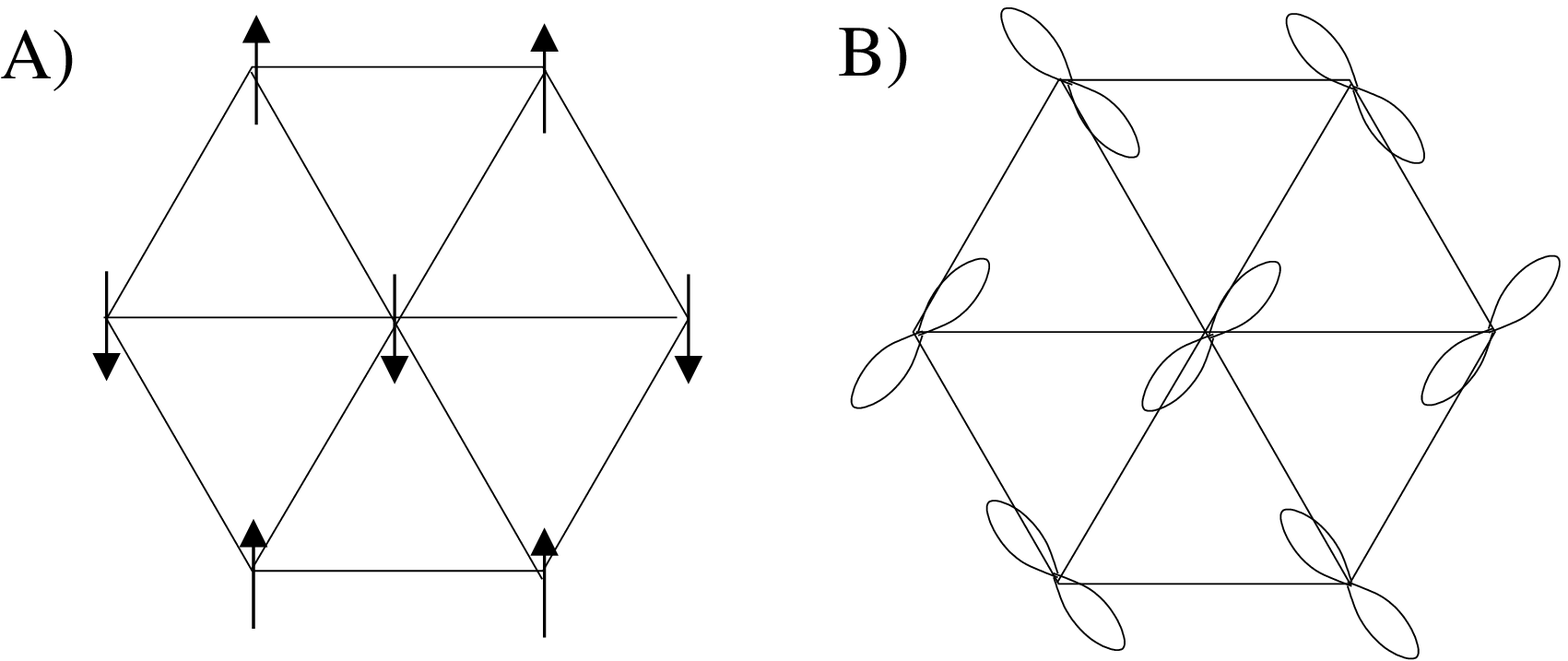,clip=1,width=0.75\linewidth,angle=0}
\vspace{3mm}
\centering\epsfig{file=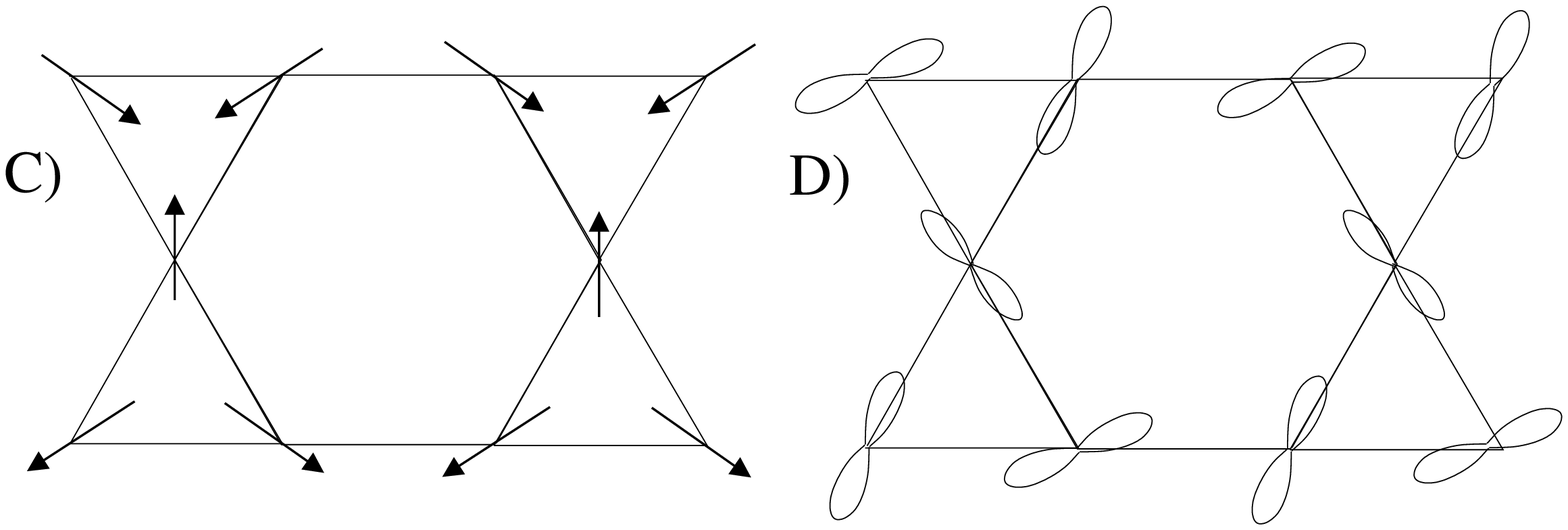,clip=1,width=0.85\linewidth,angle=0}
\caption{The $\tau$-vector (A and C) and the corresponding 
orbital (B and D) configurations.
The $p$-orbitals in the triangular lattices form 45$^\circ$ and 135$^\circ$ 
angles with the $x$-axis, and those in the Kagome lattices
are 15$^\circ$, 75$^\circ$, and 135$^\circ$.
%The triangular lattice exhibits the stripe order where
%the $p$-orbitals form the angles of 45$^\circ$ and 135$^\circ$ 
%with the $x$-axis up to a 6-fold degeneracy.
%The Kagome lattice shows the ``$Q=0$'' order where $p$-orbitals 
%form the angles of 15$^\circ$, 75$^\circ$, and 135$^\circ$ 
%angles with the $x$-axis up to a two-fold degeneracy.
}\label{fig:triag}
\end{figure}

Orbital ordering appears in all of the square, triangular, and 
Kagome lattices.
In the square lattice, Eq. \ref{eq:exchange} reduces to the 2D Ising-model
with the staggered ordering.
%on both  the
%horizontal and vertical bonds only depends on $\tau_1$, thus the 
%exchange is described by the 2D Ising-model giving rise to the staggered
%ordering of $p_x$ and $p_y$ orbitals.
For the triangular and Kagome lattices depicted in Fig. \ref{fig:triag},
we rotate $p$-orbitals on each site at $180^\circ$ around the
$x$-axis which transforms the $\tau$-vectors as 
$\tau_1\rightarrow \tau_1$ and $\tau_{2,3}\rightarrow -\tau_{2,3}$.
Correspondingly, the azimuthal angles $\varphi_\tau$ of the $\tau$-vector
and $\varphi_p$ of the $p$-orbitals satisfy
$\varphi_\tau=-2\varphi_p$ ($\varphi_p$ has a periodicity of $\pi$
instead of 2$\pi$.).
Then Eq. \ref{eq:exchange} along each bond changes to
$H_{ex}(\vec r, \vec r+\hat e_\varphi)=J_\pp
[\vec \tau (\vec r) \cdot \hat e_{\varphi}]
[\vec \tau (\vec r+ \hat e_\varphi) \cdot \hat e_{\varphi}]$
for $\hat e_\varphi=\pm\hat e_x, 
\pm(\frac{1}{2} \hat e_x \pm\frac{\sqrt 3}{2} \hat e_y)$.
In the triangular lattice, the exchange model can be reorganized into
$
H_{tri}= \frac{J_\pp}{2} \sum_{\vec r, i=1\sim 6} 
\{ [\vec \tau (\vec r) +
\vec \tau (\vec r+ \hat e_{\varphi_i})] \cdot \hat e_{\varphi_i}) \}^2
+J_\pp \sum_{\vec r} [\tau_3^2 (\vec r)-S(S+1)].
$
Thus the classic ground state configurations satisfy $\tau_3(\vec r)=0$
on each site and $[\vec\tau (\vec r)+\vec \tau (\vec r+ 
\hat e_{\varphi_i})] \cdot \hat e_{\varphi_i}=0$ on each bond.
For two neighboring sites $i$ and $j$, their azimuthal angles 
$\varphi_i$ and $\varphi_j$ of $\tau$-vectors
should satisfy either $\varphi_j=\varphi_i+\pi$ or $
\varphi_j=2\varphi(\hat e_{ij})-\varphi_j+\pi$ where $\varphi(\hat e_{ij})$ is 
the azimuthal bond angle.
It is straightforward to prove that the only classic configurations 
satisfying this constraint is depicted in Fig. \ref{fig:triag} A as 
the stripe configuration with $\tau$-vectors aligned along the 
$90^\circ$ and $270^\circ$-directions up to a 6-fold degeneracy 
associated the lattice rotation group.
The corresponding orbital configuration is shown in Fig. \ref{fig:triag} B.
For the Kagome lattice, the classical ground state configurations 
can be obtained by minimizing the exchange energy for each triangle.
It shows that $\tau$-vectors lie along the angle 
bisectors of triangles, which can be consistently arranged over
the entire lattice as the ``$Q=0$'' state depicted in
Fig. \ref{fig:triag} C and D.
Its ground state is two-fold degenerate by reversing the
directions of all of the $\tau$-vectors.
%Because the orbital exchange models in both lattices do not have
%continuous symmetry, 
Their orbital excitation spectra are gapped
%and quantum fluctuations are weak, the orbital gaps
in both the triangular and Kagome
lattices as $1.68JS$ and $2.45JS$ within a 
Holstein-Primakov type orbital wave analysis, respectively

In contrast, the $p$-orbital exchange model in the hexagonal lattice 
is markedly different, which exhibits strong orbital frustrations.
Three unit vectors  $\hat e_{1,2,3}$  denoting bond orientations
are defined as
$ \hat e_1= \hat e_x, \ \ \, \hat e_{2,3}=-\frac{1}{2} \hat e_x 
\pm \frac{\sqrt 3}{2} \hat e_y.$
Due to the bipartite nature of the honeycomb lattice, we rotate 
the $p_{x,y}$-orbitals at $180^\circ$ around the $x$-axis in 
the $A$-sublattice and around the in-plane direction of 
$\hat e_{\varphi}$ with $\varphi=45^\circ$ in the $B$-sublattice.
This transformation changes the $\tau$-operators as
$\tau_1\rightarrow \tau_1,
\tau_{2,3}\rightarrow -\tau_{2,3}$ for the $A$-sublattice and 
$\tau_{1,3}\rightarrow -\tau_{1,3}, \tau_2\rightarrow \tau_2$ 
for the $B$-sublattice.
The relations between the azimuthal angles of the $\tau$-vectors
and the $p$-orbitals are $\varphi_{\tau}=-2\varphi_{p}$ for the $A$-sublattice
and $\varphi_{\tau}= \pi-2\varphi_{p}$ for the $B$-sublattice.
We arrive at 
\bea
H_{hex} &=& J_\pp \sum_{\vec r\in A, i=1,2,3} 
( [\vec \tau (\vec r) - \vec \tau (\vec r+\hat e_i)] \cdot \hat e_i)^2 
\nonumber \\
&+&\frac{3 J_\pp}{2} \sum_{\vec r} [\tau_3^2 (\vec r)-S(S+1)].
\label{eq:120model}
\eea
A similar model is studied for the $e_{2g}$ orbitals of the 
transition metal oxides in the 3D cubic lattice \cite{biskup2005,
nussinov2004}. %, which has been shown to exhibit long range order
%in the classic limit due to an ``order by disorder'' mechanism.
%In the 2D honeycomb lattice, much stronger frustrations occur.
Eq. \ref{eq:120model} also has a similar form to the Kitaev model 
%in which the Ising quantization axes take $x,y$ and $z$ directions along 
%the three bond directions $e_{1,2,3}$,  respectively 
\cite{kitaev2006}.
In contrast, the pseudospin $\vec \tau \cdot \hat e_i$ defined here
only lies in the $xy$-plane.
%The sub-extensive $Z_2$-symmetry in the Kitaev model does not appear
%in Eq. \ref{eq:120model}.

\begin{figure}
\centering\epsfig{file=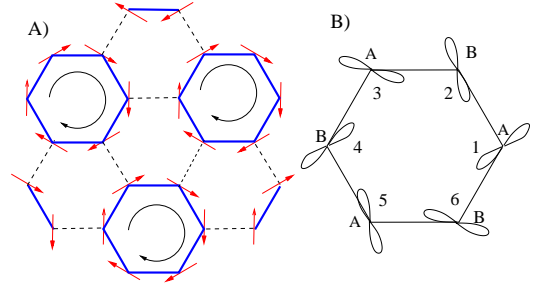,clip=1,width=0.8\linewidth,angle=0}
\caption{The fully-packed oriented loop configurations in which $\tau$-vectors
lie in directions of $\varphi=\pm 30^\circ, \pm 90^\circ, \pm 150^\circ$.
A) The closest packed loop configuration with all the loops in the same
chirality.
B) The $p$-orbital configuration for one closed loop in A).
The azimuthal angles of the $p$-orbitals are 45$^\circ$, 
105$^\circ$, 165$^\circ$, 225$^\circ$, 285$^\circ$, and 345$^\circ$. 
}\label{fig:loop}
\end{figure}

\begin{figure}
\centering\epsfig{file=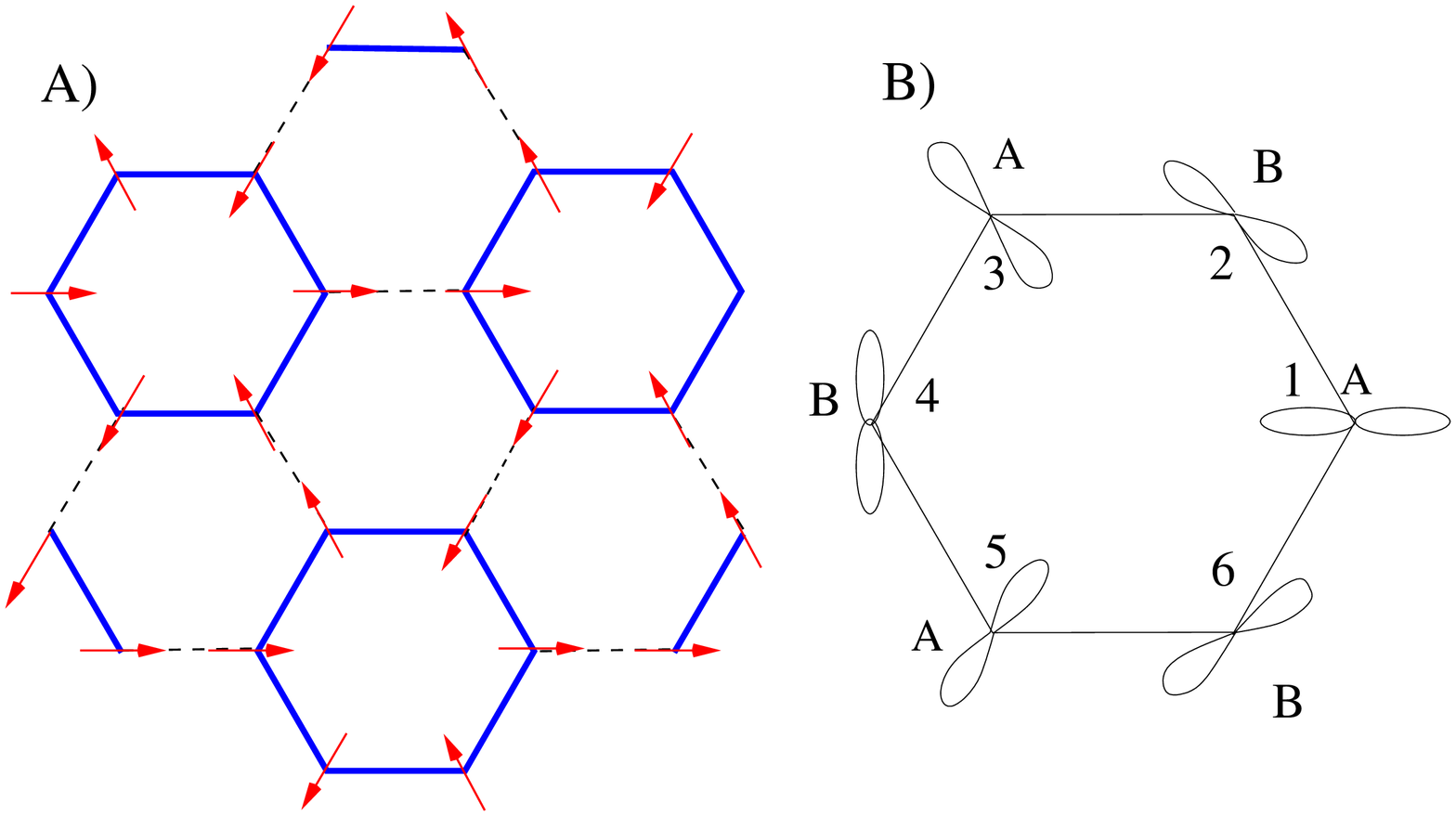,clip=1,width=0.76\linewidth,angle=0}
\centering\epsfig{file=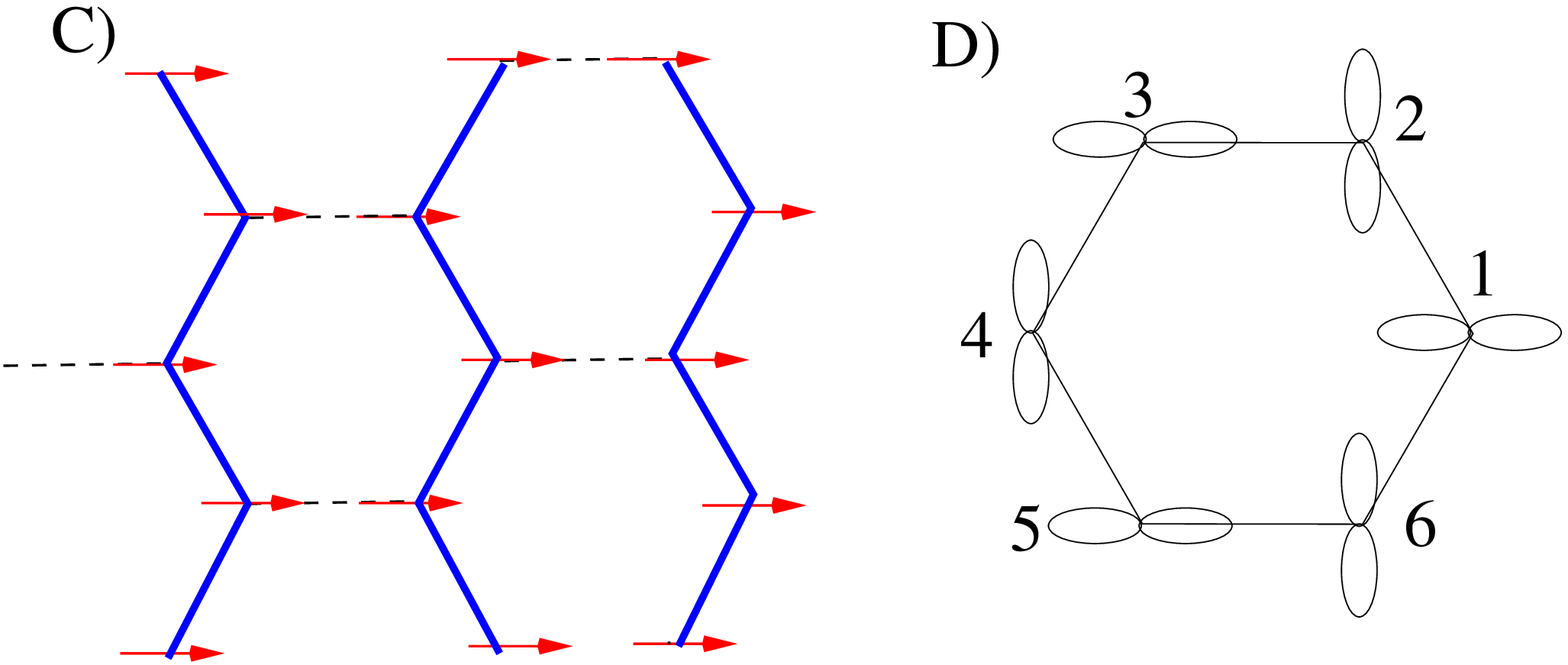,clip=1,width=0.75\linewidth,angle=0}
\caption{The fully-packed unoriented loop configurations in which
$\tau$-vectors lie along the bond directions.
A) and C) are the $\tau$-vector configurations with the closest packed loops
and the ferromagnetic state, respectively.
B) and D) are their corresponding $p$-orbital configurations.
}\label{fig:loop_2}
\end{figure}

The classic ground states of Eq. \ref{eq:120model} require that
all the $\tau$-vectors are in plane and every two $\tau$-vectors of 
a bond $\avg{ij}$ 
have the same projection along the bond direction, {\it i.e.}, 
the azimuthal angles 
$\varphi_i$ and $\varphi_j$ of the $\tau$-vectors satisfy
$\varphi(j)= \varphi(i)$, or 
$\varphi(j)= 2\varphi(\hat e_{ij}) -\varphi(i)$.
Clearly the ferromagnetic state with arbitrary in-plane polarization 
angle satisfies this constraint.
In addition, far more other classic ground state configurations can 
be constructed as follows.
Let us pick up an arbitrary lattice site $i$ and set its $\tau$-vector
angle $\varphi_i= 30^\circ$. 
Then the angle of the $\tau$-vector on any other site can only take 
one of the values of $\pm 30^\circ, 
\pm 90^\circ$, and $\pm 150^\circ$, thus it is perpendicular to 
one of the three bonds emitted from this site.
Let us  mark the rest two bonds with bold lines, then those bold lines
form loops with the $\tau$-vectors lying
tangentially to the loops.
The ground state configurations are mapped into the fully packed
non-intersecting loop configurations in the honeycomb lattice.
These loops are oriented in that the chirality of one loop
can be changed by flipping the directions of all the $\tau$-vectors 
without affecting other loops.
Fig. \ref{fig:loop} A shows one of the closest packed loop configurations 
where each loop goes around the smallest plaquette with the same chirality.
The corresponding $p$-orbital configuration along the loop
is depicted in Fig. \ref{fig:loop} B.
In the ferromagnetic states with polarization angles of 
$\pm 30^\circ, \pm 90^\circ$ and $\pm 150^\circ$, all the loops 
are infinitely long winding around the entire system.
Since the allowed loop configurations are numerous, the system
is heavily frustrated. 
It is well-known that this loop representation is equivalent to
Baxter's three-coloring model \cite{kondev1996, kondev1996a, 
baxter1970}.
If all the $\tau$-vectors are constrained to take the above
six discrete values, these allowed orientations just correspond 
to the six coloring patterns of each site in the three-coloring model.
Since each loop contains even number of bonds, we can assign two colors
(e.g. $R$ and $G$) alternatively to bonds along each loop, and the 
other one (e.g. $B$) to bonds normal to each loop.
Each loop allows two configurations (e.g. $RGRG...$ and $GRGR...$)
representing two opposite chiralities.

Next we restore the classic picture of the $\tau$-vector as a
$U(1)$ rotor in the $xy$-plane.
Each loop configuration described above has a global $U(1)$ degeneracy
associated with a suitable arrangement of the clockwise or anti-clockwise 
rotation of the $\tau$-vector on each site.
For example, for the configuration depicted in Fig. \ref{fig:loop} A,
this degeneracy corresponds to a staggered pattern of clockwise 
and anti-clockwise rotations on $\tau$-vectors in two sublattices.
For general loop configurations, the rotation directions of two
arbitrary neighboring sites are the same or opposite 
dependent on whether they have the same azimuthal angles or not.
For each six-site plaquette with arbitrary $\tau$-vector configurations, 
we have explicitly checked that rotations can be consistently arranged 
without violating the ground state energy constraint.
Since the whole lattice can be decomposed into plaquettes, rotations
can also been consistently arranged in the entire system.
If we start from one loop configuration and perform a suitable 
rotation described above at the angles of 
$n\times 60^\circ  (n=1\sim 5)$, 
we arrive at other five different oriented loop configurations. 
As a result, the classical ground state manifold of Eq. \ref{eq:120model}
is the fully-packed loop configurations multiplied by a global $U(1)$ 
rotation with the angle $-30^\circ\le \theta \le 30^\circ$.

If the rotation angle defined above is right at $30^\circ$ or 
other equivalent angles modular $60^\circ$, the $\tau$-vector
on each site is rotated to one of the three bond directions.
If we mark the other two bonds with bold lines,  they also
connect to form loops.
For example, after performing the rotation of $\pm 90^\circ$ at $A$ ($B$)
sites for the loop configuration in Fig. \ref{fig:loop} A, we arrive
at the configuration in Fig. \ref{fig:loop_2} A.
Except a global two-fold degeneracy by flipping the directions of
all the $\tau$-vectors, these loops are not oriented.
The oriented loop configurations with the same loop locations but different 
chirality distributions can be rotated into the same unoriented loop 
configuration.

\begin{figure}
\centering\epsfig{file=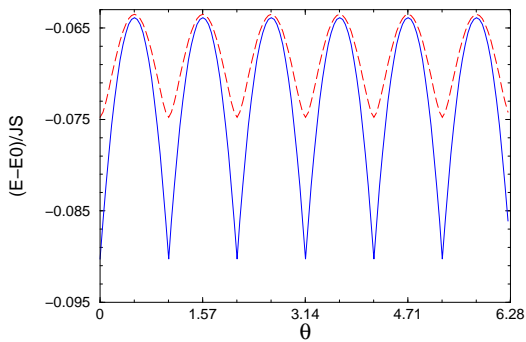,clip=1,width=0.8\linewidth,angle=0}
\caption{The ``orbital wave'' correction to the ground state energy 
per site for two manifolds of classic ground states depicted in Fig. 
\ref{fig:loop_2} A (the solid line) and C (the dashed line). 
$\theta$ is the rotation angle.
The selected ground states correspond to that depicted in Fig. 
\ref{fig:loop_2} A and its six symmetry related counterparts
in which $\theta= n \frac{\pi}{3}$ ($\theta=n\times 60^\circ$). 
}\label{fig:disorder}
\end{figure} 

So far we have elaborated the large ground state degeneracy at the
classic level, which must be lifted by quantum and thermal fluctuations.
For this purpose, we perform a Holstein-Primakov ``orbital wave'' 
analysis.
We consider the two representative ground state configurations depicted 
in Fig. \ref{fig:loop_2} A and C, and define that they correspond to 
the $\theta=0^\circ$ state in their continuous manifolds with the parameter
of the global rotation angle $\theta$.
We calculate the $1/S$-correction to the ground state energy
from the ``orbital wave'' at arbitrary angles of $\theta$.
The result is depicted in Fig. \ref{fig:disorder} and
the details will be presented elsewhere.
The ground state energies of configurations in both manifolds arrive 
at the minimum at $\theta= n\times 60^\circ (n=0\sim 5)$, i.e., 
the states represented by unoriented loops.
Furthermore, the state of Fig. \ref{fig:loop_2} A energetically wins 
over the ferromagnetic state, and such a state has an important feature: 
the appearance of the zero energy flat band of orbital modes.
This can be heuristically explained as follows.
Let us take an arbitrary six-site loop in this state.
Suppose we perform a staggered rotation with a small angle $\Delta \theta$ 
only for each site along the chosen loop but without disturbing sites in 
any other loop.
Only the six bonds connecting the chosen loop to outside increase 
energy. 
Because the $\tau$-vectors in these bonds are along the easy axis,
according to Eq. \ref{eq:120model}, the energy cost 
vanishes at the quadratic level as
\bea
\Delta E= 6 J_\pp S^2 (\Delta \theta)^4.
\eea
Each unoriented loop contributes one zero energy orbital (at the 
quadratic level).
The state of Fig. \ref{fig:loop_2} A has the
maximal number of the zero energy modes.
% and they form an entire zero 
%energy band.
As a result, the quantum zero point motion (orbital fluctuation)
selects this state as the true ground state in the large-$S$ limit.
We expect that this state not only wins over the ferromagnetic state
but also the true ground state in the large-$S$ limit, energetically 
better than any other configuration which always has less number
of zero energy orbital modes.
Experimentally, this ordering pattern with the enlarged unit cell of six sites
can be easily detected in the time of flight noise correlation spectra. The
second-order coherence peaks will appear at the reciprocal lattice vectors
of the corresponding reduced Brillouin zone.

Next we briefly discuss the effects from the $t_\perp$ term and finite 
temperatures.
The $t_\perp$ term generates the orbital flipping process as
$ 
\Delta H(\vec r, \vec r^\prime ) 
= J_\perp \big\{ -(\vec \tau (\vec r) \cdot 
\hat e_{\vec r \vec r^\prime}^\prime )
(\vec \tau (\vec r^\prime) \cdot \hat e^\prime_{\vec r \vec r^\prime}) 
+ \tau_3 (\vec r) \tau_3 (\vec r^\prime) \big \},
$
where $\hat e^\prime_{\vec r \vec r^\prime}$ lies in plane and is 
perpendicular to $\hat e_{\vec r \vec r^\prime}$.
For the two ground state configurations depicted in Fig. \ref{fig:loop_2} 
A and C, this term favors the ferromagnetic state at the classic level
by gaining the energy of $\Delta E_{cl}=J_\perp S^2$, but pays the cost
of the zero point fluctuation energy around $\Delta E_{flc}=0.01 J_\pp S$
as shown in Fig. \ref{fig:disorder}. 
Due to the smallness of the $J_\perp/J_\pp=t_\perp/t_\pp$,
$\Delta E_{flc}$ and $\Delta E_{cl}$ are close to each other and
lead to rich phase competitions.
For the realistic system where $S=\frac{1}{2}$, the plaquette phase
in Fig. \ref{fig:loop_2} c is stabilized roughly at $t_\perp<0.01 t_\pp$
which can be easily realized in the realistic system as calculated 
in Ref. \cite{wu2007a}.
On the other hand, thermal fluctuations also help to stabilize the
plaquette state which has the maximal number of zero modes by enhancing
the entropy contribution.

In summary, we have presented the general structure of the $p$-orbital
exchange physics, which gives rise to many different features
from the $d$-orbital solid state systems,
including the orbital ordering in triangular and Kagome lattices and orbital 
frustration of the 120$^\circ$ orbital model in the honeycomb lattice.
The six-site plaquette ordering pattern in the honeycomb 
lattice is found due to the ``order from disorder'' mechanism.
Although the above analysis was done at the large-$S$ level, 
it is well-known that quantum fluctuations at 2D usually are not
strong enough to destroy long range order.
It is conceivable that the above orbital orderings also extrapolate 
to the real orbital systems at $S=1/2$.
For example, spin orderings in square and triangular lattices of 
quantum magnets by large-$S$ methods also apply to the spin-$1/2$ case.

%Many questions remain open for future exploration.
%In particular, the ground state of the 120$^\circ$-orbital model
%in the quantum limit, i.e., $s=\frac{1}{2}$, is a challenge problem
%and provide a promising direction to realize the exotic quantum
%disordered orbital liquid state.
%The $p$-orbital Mott-insulators with spin degree of freedom provide
%a new system to study the competition between orbital and spin
%\cite{wu2008}.
%Finally, whether the exotic quantum disordered orbital liquid state 
%can be identified in optical lattices is a most intriguing 
%problem. 

%Moreover, I have also found orbital frustration in the 3D cubic 
%lattice system and in $p$-orbital bosonic Mott-insulators.
%Another important question for future research is how to
%experimentally detect the possible orbital liquid state in
%optical lattices.

C. W. thanks C. Nayak, Z. Nussinov for helpful discussions.
C. W. is supported by the start up funding at UCSD
and Sloan Research fellowship.
{\it Note added}\ \ \
Upon the completion of this manuscript, there appeared an independent work  
by E. Zhao {\it et al.} \cite{zhao2008} on the orbital exchange model but
with different conclusions.
%However, many conclusions in our paper are different from theirs.

%\begin{acknowledgements}
%\end{acknowledgements}
%\bibliographystyle{prsty}
%\bibliography{orbital}

\end{document}